\documentclass[12pt]{article}
\usepackage{amsfonts,amssymb,amsbsy}
\usepackage{graphicx}% Include figure files
\def\pacs{\noindent PACS:~}
\begin{document}
\title{Quantum gravity as a consistent local field theory}
\date{}
%\vspace*{5mm}
\author{V.V.Kiselev,\\~ \normalsize
State Research Center "Institute for High Energy Physics" \\
{~ \normalsize Protvino, Moscow region, 142280 Russia}\\
~ \normalsize Fax: +7-0967-744937, E-mail: kiselev@th1.ihep.su
}

\maketitle

\begin{abstract}
We show that the Einstein-Hilbert action for the gravitational field can be
obtained as a linear low-energy approximation for the dynamical massless fields
in the theory with the lagrangian quadratic in the gauge field strength-tensor
of spin connection under the spontaneous breaking of symmetry with a vacuum
state described by an ansatz for nontrivial background strength-tension.
\end{abstract}

\section{Introduction}
There is the only physical argument for the development of string, superstring
and M-theories with extended objects in spaces with extra dimensions as well as
some other popular topics of mathematical physics \cite{M}. The motivation is
the problem on a consistent local quantum field theory of gravitational
interactions, since, under the quantization, the classic Einstein--Hilbert
action for the gravitational field leads to the nonrenormalizable theory. The
nonrenormalizability implies that the quantum loop corrections cannot be
calculated in terms of finite number of physical parameters determining the
coupling constant of interaction and the normalization of fields. Instead of
that, some infinite set of constants appears in the series of operators with
growing powers of energy. In this way, a small contribution of such corrections
as observed empirically, can be explained by the fact that the standard
gravitational interaction is the particular contribution giving the valid
description in the restricted region of energy, only. This contribution is
originated from a consistent full theory, so that the matching of full theory
parameters with the couplings of higher operators results in the suppression of
higher operators. The physical reason for the nonrenormalizability is caused by
the dimension of gravitational constant determining the interaction of
dynamical metric field with the energy-momentum tensor. The presence of
fundamental dimensional parameter can be naturally introduced\footnote{In the
renormalizable gauge theories with local, point-like, fields the dimensional
quantities appear due to the spontaneous breaking of symmetry or in the
dimensional transmutation in the renormalization group, but in the bare
lagrangians.} in the theories with the extended objects like the strings,
membranes and so on, so that the consistent quantization of such the
objects\footnote{We mean the quantization with no anomalies.} can be done only
in the spaces of higher dimensions. In this way, the principles of
renormalizability and gauge invariance lose their fundamental role in the
construction of action and become auxiliary in the reduction of the theory to
the observed four-dimensional world. Thus, the quantum gravity makes the
question on the principal consistency and completeness of local field theory.

In this paper we develop the idea by Utiyama on the gauge invariance of
gravitational interactions \cite{Utiyama} and show how the action quadratic in
the strength tensor is reduced to the standard Einstein--Hilbert action with
the massless gravitons in the low-energy limit under the spontaneous breaking
of symmetry by introducing an ansatz for the vacuum state. As was
demonstrated by Stelle \cite{stelle}, the introduction of terms quadratic in
the Riemannian tensor of curvature provides the renormalizability of quantum
theory, that is rather natural for the general nonabelian gauge theory.
Therefore, we get the real possibility to solve the problem on the formulation
of quantum gravity in the framework of local field theory.

\section{Spontaneous breaking of symmetry}
The starting point of consideration is the action of spin-connection
gauge-field $\cal A$ interacting with the spinor $\theta$,
\begin{eqnarray}
S &=& \int {\rm det}[h]\, d^4x\; \left\{\frac{1}{4 g^2_{\rm\scriptscriptstyle
Pl}}\, {\cal F}_{\mu\nu, mn} {\cal F}^{\mu\nu, mn} + 
%\right.\nonumber \\ &+& \left.
\frac{\rm i}{2} [\bar
\theta(x)\, \bar \sigma^{\mu}\nabla_\mu \theta(x) + 
\theta(x)\, \sigma^{\mu}\overline\nabla_\mu \bar \theta(x) ]\right\},
\label{ls-action}
\end{eqnarray}
where we accept the usual notations in the textbook by Wess and
Bagger \cite{BW}, $h^\mu_m$ is a tetrad, $g_{\rm\scriptscriptstyle Pl}$ is a
coupling constant of the gauge interaction. The lagrangian in (\ref{ls-action})
is invariant under the local gauge transformations by the group SL$(2,\mathbb
C)$ acting on the Weyl spinors
\begin{equation}
\begin{array}{rclrcl}
\theta^\prime &=& M \theta, & \bar\theta^\prime &=& \bar \theta M^\dagger,
\\
\sigma^{\prime\mu} &=& M\, \sigma^\mu \, M^\dagger, & 
\bar\sigma^{\prime\mu} &=& [M^\dagger]^{-1}\, \bar\sigma^\mu \, M^{-1},
\end{array}
\end{equation}
whereas the infinitesimal transformations with the parameters $\omega_{mn}$
\begin{equation}
\begin{array}{rcl}
M = 1 + \sigma^{nm} \omega_{nm},\\[2mm]
[M^\dagger]^{-1} = 1 + \bar\sigma^{nm} \omega_{nm},
\end{array}
\quad \omega_{nm}\to 0,
\label{inf}
\end{equation}
are given by the generators
\begin{equation}
\begin{array}{rcl}
[\sigma^{nm}]_{\alpha}^{~\beta} &=& \frac{1}{4}\left[
\sigma^{n}_{\alpha\dot\alpha}\bar\sigma^{m\dot\alpha\beta} - 
\sigma^{m}_{\alpha\dot\alpha}\bar\sigma^{n\dot\alpha\beta}\right],\\[2mm]
[\bar\sigma^{nm}]^{\dot\alpha}_{~\dot\beta} &=& \frac{1}{4}\left[
\bar\sigma^{n\dot\alpha\beta}\sigma^{m}_{\beta\dot\beta} - 
\bar\sigma^{m\dot\alpha\beta}\sigma^{n}_{\beta\dot\beta}\right].
\end{array}
\end{equation}
which satisfy the ordinary commutation relations for the spin operators. The
covariant derivatives on the left-handed and right-handed chiral spinors
\begin{equation}
\begin{array}{rcl}
\nabla_{\mu\alpha}^{~~~\beta} &=& \delta_\alpha^{~\beta}\, \partial_\mu + 
{\cal A}_{\mu,nm} \sigma^{nm~\beta}_{~~\alpha}, \\[2mm]
\overline\nabla_{\mu~\dot\beta}^{~\dot\alpha} &=&
\delta^{\dot\alpha}_{~\dot\beta}\, \partial_\mu +  {\cal A}_{\mu,nm}
\bar\sigma^{nm\dot\alpha}_{~~~~~\dot\beta},
\end{array}
\end{equation}
determine the strength tensor of gauge field ${\cal A}$
\begin{equation}
\begin{array}{rcl}
[\nabla_\mu,\, \nabla_\nu] &=& {\cal F}_{\mu\nu, mn}\, \sigma^{mn},\\[2mm]
[\overline\nabla_\mu,\, \overline\nabla_\nu] &=& {\cal F}_{\mu\nu, mn}\,
\bar\sigma^{mn},
\end{array}
\label{tensor}
\end{equation}
so that 
\begin{equation}
\begin{array}{rcl}
{\cal F}_{\mu\nu, mn} &=& \partial_\mu {\cal A}_{\nu,mn} - \partial_\nu {\cal
A}_{\mu,mn} + 
%\\[2mm] &+& 
2 ({\cal A}_{\mu,mk} {\cal A}_{\nu,ln} - {\cal A}_{\nu,mk} {\cal
A}_{\mu,ln})\, g^{kl}.
\end{array}
\label{strength}
\end{equation}
The Noether current, which follows from the gauge invariance under the change
of equivalent representations of spinor algebra, has the form 
\begin{equation}
j^{\mu,nm} = \frac{\rm i}{2} [\bar \theta(x)\, \bar \sigma^{\mu}\, \sigma^{nm}
\theta(x) + 
\theta(x)\, \sigma^{\mu}\, \bar \sigma^{nm}\bar \theta(x) ].
\label{current}
\end{equation}
In the Pauli gauge for the $\sigma$-matrices, $\sigma^{\mu} = (1,{\boldsymbol
\sigma})$, we get
\begin{equation}
j^{\mu,nm} = \frac{1}{2}\, \epsilon^{\mu nm \nu}\, j^{\lambda} \,
g_{\nu\lambda},\quad 
j^{\lambda} = \bar \theta\, \bar\sigma^{\lambda}\,\theta.
\label{currents}
\end{equation}
Further, in the usual way we can derive the current conservation and the
Slavnov--Taylor identities (see discussions in the extended version of
\cite{weyl}).

Introduce some nonzero vacuum fields leading to the spontaneous breaking of
symmetry, so that
\begin{equation}
\begin{array}{rcl}
{\cal F}_{\mu\nu, mn} &=& {\cal R}^{\rm vac}_{\mu\nu, mn} + 
%\\[2mm] &+& 
\frac{1}{2}{\cal
R}_{\mu\nu, mn}[\Gamma]-\frac{1}{2}\, \epsilon_{\mu\nu}^{~~\alpha\beta}
\,\frac{1}{2}{\cal R}_{\alpha\beta, mn}[\Gamma^D],
\end{array}
\label{vacF}
\end{equation}
where we suppose the following ansatz:
\begin{equation}
\begin{array}{rcl}
{\cal R}^{\rm vac}_{\mu\nu, mn} &=& -  (g_{\mu m}g_{\nu n} - 
g_{\mu n}g_{\nu m})\, g^2_{\rm\scriptscriptstyle Pl} v^2 - 
%\\[2mm]&& -
\epsilon_{\mu\nu mn} g^2_{\rm\scriptscriptstyle Pl}v^2,
\end{array}
\label{vevF}
\end{equation}
while $\Gamma$ and $\Gamma^D$ are dynamical fields redefined in accordance with
$\Gamma_{\mu,nm} = {2}{\cal A}_{\mu,nm}$, and ${\cal F}_{\mu\nu,
mn}[{\cal A}] = \frac{1}{2} {\cal R}_{\mu\nu,nm}[\Gamma]$, so that ${\cal
R}_{\mu\nu,nm}[\Gamma]$ represents the Riemannian tensor of curvature as we
will see below. Then the contribution into the lagrangian linear in the vacuum
fields\footnote{We remove the terms like $v^2 \epsilon^{\mu\nu mn}{\cal
R}_{\mu\nu mn}$, since they are equal to zero with the symmetric connection in
the world indices.}, takes the form
\begin{equation}
{\cal L}_G =  - \frac{1}{2}\,v^2\; {\cal R}[\Gamma] - \frac{1}{2}\,v^2\; {\cal
R}[\Gamma^D],
\label{gravityd0}
\end{equation}  
where we have introduced the Ricci tensor ${\cal R}_{\mu\nu}[\Gamma] = {\cal
R}^{\gamma}_{~~\mu,\gamma\nu}[\Gamma]$ and the scalar curvature
${\cal R}[\Gamma] = {\cal R}_{\mu\nu}\, g^{\mu\nu}$. Following Palatini
\cite{pala}, we find that the variation of action over the connection gives
zero covariant derivative of metric, i.e. it leads to the metric connection
expressed in terms of Cristoffel symbols for both $\Gamma$ and $\Gamma^D$, and
they are coherent $\Gamma=\Gamma^D$. The variation over the metric gives the
Einstein equations.

The lagrangian of (\ref{gravityd0}) coincides with the Einstein--Hilbert
lagrangian of general relativity, if we put
$
v^2 = \frac{1}{16\pi G},
$
where $G$ is the gravitational constant, and then 
\begin{equation}
{\cal L}_G =  - \frac{1}{16\pi G}\; {\cal R}[\Gamma].
\label{gravityd}
\end{equation}
Let us stress the most important features of suggested mechanism for the
spontaneous breaking of gauge symmetry:
\begin{enumerate}
\item
The vacuum expectation value of (\ref{vevF}) is covariant, and it preserves the
symmetry with respect to general transformations of coordinates.
\item
The expression of (\ref{vevF}) represents the sum of the Riemannian tensor for
the space-time with the constant curvature (the first term) and that of dual
one.
\item
The field of (\ref{vevF}) can satisfy the gauge-field equations at zero
external sources, if the covariant derivative of metric is equal to zero.
\item
The cosmological constant caused by the term quadratic in the vacuum
strength-tensor, is equal to zero.
\end{enumerate}

\section{Conclusion}
In this paper we have considered an ansatz for the vacuum state leading to the
spontaneous breaking of gauge symmetry in the theory with the spin connection.
To the linear approximation in the background vacuum fields this mechanism
leads to the Einstein--Hilbert action for the gravitons. Therefore, we find the
opportunity to solve the problem on the formulation of quantum gravity in the
framework of local field theory.

If there are no anomalies, the gauge invariance insures the renormalizability
of the theory. However, we have to point out the necessity of investigations
concerning for massive modes appearing under the spontaneous breaking of
symmetry, the canonical quantization and the role of duality, which is
discussed in the extended version \cite{weyl}. In addition, the nonabelian
character of gauge group leads to the asymptotic freedom of running coupling
constant in the renormalization group. Thus, the principal idea on the quantum
gravity supposed in this paper demands a further consideration.

This work is in part supported by the Russian Foundation for Basic Research,
grants 01-02-99315, 01-02-16585 and 00-15-96645.

%\newpage 

\end{document}